\documentclass[twocol]{ametsocV6.1}

\usepackage{multirow} 

\title{Global Kilometer-Scale Simulations with ARP-GEM2: Effect of Parameterized Convection and Calibration}

\authors{Olivier GEOFFROY\aff{a}\correspondingauthor{Olivier Geoffroy, olivier.geoffroy@meteo.fr} and David SAINT-MARTIN\aff{a}}
\affiliation{\aff{a}{CNRM, Université de Toulouse, Météo-France, CNRS, Toulouse, France}}

\abstract{The objective of this paper is twofold. First, it documents the second version of the global atmospheric model ARP-GEM and its calibration at kilometer-scale resolution. The model is currently able to run simulations at a resolution of up to 1.3 km. Second, this paper focuses on multi-year global atmospheric simulations at a 2.6 km resolution with and without parameterized convection and associated calibration. Simulations without deep convection tend to be similar to those with infinite, or at least large, entrainment values. Consistently, entrainment and detrainment can be used as primary drivers for the gradual reduction of convection as resolution increases. The results indicate that, in this hydrostatic model, parameterized convection still plays a significant role in the correct representation of the mean state at the kilometer scale. Additionally, they suggest some added value of high resolution in representing climate variability. However, a compromise between the adequate representation of the mean state and variability is necessary, as both are differently favored by the degree of parameterized convection. 
}

\begin{document}

\maketitle

\section{Introduction}

Climate models have evolved only modestly across successive Coupled Model Intercomparison Project (CMIP) exercises, and many systematic biases persist from one generation of climate model to the next \citep[e.g.][]{bock-2020, tian-2020, ahn-2023}. The subgrid-scale representation of moist atmospheric processes plays a key role in shaping these biases and overall model behavior \cite[e.g.][]{stevens-2013}. The emergence of kilometer-scale modeling may represent a significant advance to improve the representation of convective processes -- or at least a necessary step toward higher spatial resolution. As models reach this finer scale, the need for convective parameterization schemes progressively decreases. This motivates the investigation of the scientific and practical implications of higher-resolution models.

In recent years, simulations of one year or more at resolutions below 5 km have become feasible, and their use is becoming increasingly widespread \citep[e.g.][]{satoh-2005, satoh-2008, hohenegger-2009, hohenegger-2020, wedi-2014, stevens-2019, satoh-2019, tomita-2005, freitas-2020, caldwell-2021, giorgetta-2022, donahue-2024, takasuka-2024a, takasuka-2024b, dipankar-2026}. In addition to standard AMIP-type experiments, these developments include climate change experiments with prescribed sea surface temperature increases \citep{tsushima-2014, merlis-2024}, multi-year atmosphere–ocean coupled experiments \citep{rackow-2025}, and shorter, ultra-high-resolution experiments with grid spacings approaching 1 km \citep{wedi-2020, fuhrer-2018}.
Until now, calibration of these models has been challenging due to their high computational cost, which may limit their practicality compared with CMIP-class models \citep{schneider-2024}.

Convection encompasses a wide range of spatial and temporal scales, from individual convective cells of varying size to large, organized clusters of convection. As the model resolution approaches the large-eddy scale, subgrid-scale convective fluxes from the convection schemes should vanish. This reduction in convective fluxes should occur gradually with increasing resolution to ensure a smooth transition from parameterized to explicitly resolved convection \citep[e.g.][]{arakawa-2011, malardel-2019}.

Currently, the degree of complexity of kilometer-resolution atmospheric models varies \citep[e.g.][]{satoh-2019}. Some simulations still use a hydrostatic core \citep{wedi-2020, rackow-2025}, and models display a wide diversity in parameterizations, from simplified physics typical of LES models (no convection, Smagorinsky-type turbulence, “all-or-nothing” condensation) to more complex physics typical of climate models \citep[e.g.][]{stevens-2019}.

At resolution lower than 10 km, some features of the system -- such as tropical variability or the diurnal cycle -- may actually be better represented without deep convection parameterization \citep{sato-2009, miura-2007, vergara-temprado-2020, takasuka-2026}. While convection schemes are necessary to overcome some model limitations, they can also distort certain aspects of the simulation due to the inherent approximations involved. 
Moreover, the many effects introduced by these parameterizations and their interactions complicate model understanding.

On the other hand, some aspects of the climate are improved with convection, and the need for convection parameterization, to some extent, remains true even at the kilometer scale \citep[e.g.][]{freitas-2020, arnold-2020}. Specifically, convective parameterizations can help mitigate certain model biases and improve the representation of convective systems and precipitation characteristics, particularly the excessive intensity of heavy precipitation events \citep{ becker-2021, takasuka-2026}.
Large eddy simulations over limited-area domains indicate that intermediate regimes, such as congestus clouds, still need to be parameterized \citep{champouillon-2023}.
This is particularly true given that the model’s effective resolution -- the smallest scale the model effectively resolves -- is 4 to 10 times larger than the nominal grid spacing, with largest differences for semi-implicit semi-Lagrangian models \citep{Abdalla-2013,ricard-2013}. 
Consequently, even at 1-2 km grid spacing, a significant portion of the convective energy spectrum remains parameterized, with roughly half of the cumulative convective energy still below the resolved scales at resolutions lower than 1.5 km \citep{schneider-2024}.

In this context, we have developed the global, efficient and multi-resolution atmospheric model ARP-GEM, a highly optimized version of ARPEGE/IFS and presented a suite of simulations up to 6 km resolution \citep[][hereafter GS25]{geoffroy-2025}. The present paper documents version 2 of the ARP-GEM model, along with a set of kilometer-scale simulations. This version is also used within the context of the third phase of the DYAMOND intercomparison project \citep{takasuka-2024b}, at resolutions of 2.6 km and 1.3 km. 
We present multi-year global atmospheric simulations at 2.6 km resolution, with and without parameterized convection, and discuss their calibration. Special attention is given to deep convection, which requires reduced intensity at higher resolutions

Section 2 presents ARP-GEM version 2 and kilometer-scale configurations, Section 3 describes the model’s computational efficiency at resolutions up to 1.3 km, Section 4 details the simulation setups and addresses resolution-dependent tuning, while Section 5 discusses the sensitivity to the convection scheme at 2.6 km resolution and compares the results with coarser resolutions.

\section{The ARP-GEM2 atmospheric model}
\label{sec:model}

\subsection{Model Description Overview}

This section presents the Global Efficient and Multiscale (ARP-GEM) atmospheric model version 2, hereafter referred to as ARP-GEM2. It is a revised version of ARP-GEM1, which is described in detail in GS25.
The model employs a semi-implicit, semi-Lagrangian spectral dynamical core, which allows for large timesteps. It is formulated under the hydrostatic assumption. 
The grid-scale cloud scheme is based on \cite{smith-1990} and assumes a subgrid PDF of thermodynamic variables with a fixed vertical profile of critical relative humidity.
The grid-scale microphysics scheme \citep{lopez-2002} is a minimal one-moment bulk scheme with four species (cloud liquid, cloud ice, precipitating liquid, and precipitating solid), using Kessler-type autoconversion rates \citep{kessler-1969, lin-1983} and assuming constant vertical velocities for sedimentation.
The model uses a level 2.5 TKE turbulence scheme based on \cite{cuxart-2000}, with a non-local mixing length following \cite{bougeault-1989} and extended in GS25 to include moist processes; a shallow convection scheme representing dry and moist thermals with a surface-flux closure (GS25); and a deep convection scheme based on the bulk updraft formulation of \cite{tiedtke-1989}.

The modifications introduced in ARP-GEM2 are specific to high-resolution modeling and aimed to improve the model's physical performance and consistency. An overview of the changes is provided here, with additional technical details presented in Appendices.

\subsection{Physics updates}
\label{sec:physics}
This section describes physical modifications and tuning adjustments made in transitioning from ARP-GEM1 to ARP-GEM2. The resolution-dependent tuning, particularly the treatment of deep convection, is addressed in Sections \ref{sec:simulations}\ref{sec:diluted},\ref{sec:resoltun}.
Most physics modifications concern convection, particularly shallow convection, and associated recalibration, along with modifications in the general tuning of the model. These modifications aimed at reducing some biases observed in GS25, ensuring consistency with other schemes, adapting to high-resolution modeling, and including some design adjustments and modeling choices that have limited impact on overall simulation outcomes. These modifications are described in more detail in the Appendix.

The deep convection triggering has been revised, reverting to the formulation originally proposed by \citet{jakob-2003} and  
the intensity of deep convection is reduced (Appendix \ref{app:deep}).
The degree of dilution, through adjustment of entrainment and detrainment rates, is used as a main tuning parameter (Section \ref{sec:simulations}\ref{sec:diluted}).

Concerning boundary layer clouds, a significant improvement is better modulation of mixing at the top of the boundary layer to prevent the destruction of stratocumulus cloud coverage. 
In ARP-GEM2, mixing at the top of the cloudy boundary layer is restricted to turbulent entrainment by preventing shallow mixing where grid-scale clouds occur at the inversion (see Appendix \ref{app:topentr}). This approach prevents an overlap in the representation of the mixing process and promotes a better representation of stratocumulus layers.
Note, however, that low-level cloud cover decreases with increasing resolution. Therefore, the increase in stratocumulus, which is related to reduced mixing, is modulated by this resolution effect at kilometer-scale resolution.

Further developments in the physics directly concern the shallow convection scheme. 
In general, the modifications to the shallow convection scheme make it less intense and shallower in terms of precipitation and transport. 
Particularly high precipitation at high resolution in GS25 was partly related to shallow convection. Other changes also aim to enhance physical realism, to the extent that a mass flux scheme can be considered physically realistic.
These additional modifications are described in Appendix \ref{app:shall} and do not appear to have a strong impact on the model simulations.
They include revisions in shallow precipitation and in-cloud water content formulations, detrainment and entrainment rates, momentum transport, detrainment of turbulent kinetic energy (TKE), the inclusion of a maximum depth in the shallow cloud definition, and the initial properties of updrafts. 
Entrainment is globally increased, consistent with higher-resolution simulations, for which convection should be globally less diluted.
Unlike deep convection, which is may be adjusted for each resolution, the shallow convection parameters remain fixed once and for all, for simplicity. Note that shallow and deep convection should ultimately be represented by a single, unified scheme, for consistency.

The intensity of turbulent mixing is increased by decreasing the TKE dissipation (Appendix \ref{app:turb}). 
This aimed to limit the predominant cold bias in surface air temperature.
Finally, some parameters in the microphysics are modified (Appendix \ref{app:micro}).

\begin{table*}[h]
\caption{Configuration details and computational performance for the four simulations. The grid o$N_g$ refers to a octahedral reduced Gaussian grid with $N_g$ Gaussian latitudes and 2$N_g$ longitudes along equatorial Gaussian latitudes. The coarsening factor refers to the ratio between the grid-point model resolution and the radiative grid resolution. SDPD refers to as Simulated Days Per Day.}
\begin{center}
\begin{tabular}{ccccccccc}
\topline
Configuration & Grid Point & Spectral & Time & Rad. \& Surf. & Coarsening & Radiation & CPU & SDPD \\
Name & Resolution (km) & Truncation & Step (s) & Resolution & Factor & Timestep (s) & Cores &  \\
\midline
ARP-GEM2-25km & o782 (25 km)  & 390  & 900 & o244  (82 km) & 3.2 & 7200 & 9x128 & 6000 \\
ARP-GEM2-12km & o1564 (12.6 km)& 781  & 600 & o488  (41 km) & 3.2 & 3600 & 26x128 & 2500 \\
ARP-GEM2-2.6km  & o7680 (2.6 km)  & 3839 & 240 & o1310 (15 km) & 5.8 & 1200 & 145x128 & 173 \\
ARP-GEM2-1.3km  & o15360 (1.3 km) & 7679 & 120 & o1310 (15 km) & 11.6 & 1800 & 361x128 & 46 \\
\botline
\end{tabular}
\label{tab:configurations}
\end{center}
\end{table*}

\subsection{Configurations at the kilometer scale}

The ARP-GEM model has undergone new developments to support atmospheric global simulations at horizontal resolutions of up to 1.3 km. 
In particular, the possibility of using double precision has been reintroduced in large portions of the spectral transformations, including the fast Legendre transform and the fast Fourier transform. This option is activated in the 1.3 km configuration to mitigate the occurrence of spurious “lined” or “striped” patterns that may appear in some physical fields, such as precipitation. 
Additional numerical inaccuracies linked to single precision are addressed by using double precision in some parts of the semi-Lagrangian advection, such as the calculation of the longitude and latitude of interpolation points.
Other technical developments and changes in dynamics options are detailed in Appendix \ref{app:kmscale}.

Given the limited vertical resolution in ARP-GEM1, the number of vertical levels has been increased from 50 to 72 in ARP-GEM2.
The 72-vertical-level grid is also documented in GS25. It is a low-top model with a small number of vertical levels in the stratosphere, thereby concentrating vertical levels in the troposphere.
The vertical grid spacing increases with altitude, reaching roughly 400 m in the free troposphere.
This value is consistent with the maximum value recommended by  \cite{schmidt-2024}, but much larger than the 200 m vertical grid spacing for which \cite{skamarock-2019} find convergence of their results.
The change from 50 to 72 levels helps reduce biases in the extratropical upper troposphere that appear at higher horizontal resolutions, as seen in the 6 km ARP-GEM1 configuration (e.g., Fig. 14 in GS25) or in the Nonhydrostatic ICosahedral Atmospheric Model NICAM \citep[e.g. Fig. 13 in][]{takasuka-2024a}.
The revised vertical grid is used consistently across all horizontal resolutions in ARP-GEM2.

Two configurations at the kilometer scale have been developed. Their specificities are summarized in Table \ref{tab:configurations}, along with the 25 km and 12.6 km configurations also used in this study for comparison. The 25 km and 12.6 km configurations share the same characteristics as those described in GS25 with 72 vertical levels. 

The 2.6 km configuration uses a TCo3839 grid (where T stands for truncation, C for cubic, and o for octahedral), corresponding to a spectral truncation of $n=3839$ and an octahedral reduced grid with $N_{g} = 7680$ Gaussian latitudes and $15360$ longitudes along the equatorial Gaussian latitudes. The 1.3 km configuration uses a TCo7679 grid, corresponding to a spectral truncation of $n=7679$ and an octahedral reduced grid with $N_{g} = 15360$ Gaussian latitudes and $30720$ longitudes along the equatorial Gaussian latitudes. The semi-implicit, semi-Lagrangian formulation of the dynamical core, combined with a physics compatible with long time steps, allows the use of large time steps of 240 s and 120 s for the 2.6 km and 1.3 km configurations, respectively.

Temporal coarsening is applied to radiation calculations, and spatial coarsening is used for both radiation and surface flux calculations, while topography is maintained at atmospheric resolution. At a horizontal resolution of 2.6 km, the radiative timestep is 20 minutes, corresponding to one radiation call every five model time steps. The spatial coarsening factor is 5.8, meaning that the surface and radiation grids are 5.8 times coarser than the atmospheric grid.
In the 1.3 km configuration, the radiation time step is larger than that of the 2.6 km configuration. The 1.3 km simulation presented here is an older exploratory run, with specificities that have not yet been revised since the first tests, unlike the 2.6 km simulation. The 1.3 km spatial coarsening factor (11.6) is much larger than that of the 2.6 km simulation (5.8) because it uses the same grid for coarsening as the 2.6 km configuration.

Using a coarsened grid for radiation and surface fluxes is consistent with the fact that the effective resolution for atmospheric motion is significantly coarser than that of the atmospheric grid, while these processes are assumed to be one-dimensional in the model. In addition, the validity of the column hypothesis for the surface model -- assuming no horizontal exchanges -- may be compromised at resolutions finer than 10 km.
Low-level clouds remain parameterized at this resolution, which necessarily limits the accuracy of their representation and their coupling with the large-scale circulation. This reduces the added value of a high-fidelity representation of cloud–radiation interactions. Additionally, sensitivity tests show only a small impact on stratocumulus decks when radiation is applied homogeneously over an entire LES domain \citep{bellon-2016}. 
Concerning deep convection, latent heat release is expected to dominate its dynamics, and anvil clouds cover large areas, which tends to make a coarsened grid acceptable for representing their radiative effect.

To gain more insight into the role of temporal and spatial coarsening, we perform two additional two-year simulations at 2.6 km resolution, alongside the simulations presented in Section \ref{sec:simulations}\ref{sec:resoltun}. In the first, the radiation timestep is set equal to the model timestep (4 min instead of 20 min). In the second simulation, the spatial coarsening factor is reduced from 5.8 to 3, corresponding to a coarsened grid with a resolution of approximately 7.8 km. The results, presented in the Supplementary Material, show no significant sensitivity.

\subsection{Computational performance at high resolution}
\label{sec:performances}

\begin{figure}
\centerline{\includegraphics[width=19pc]{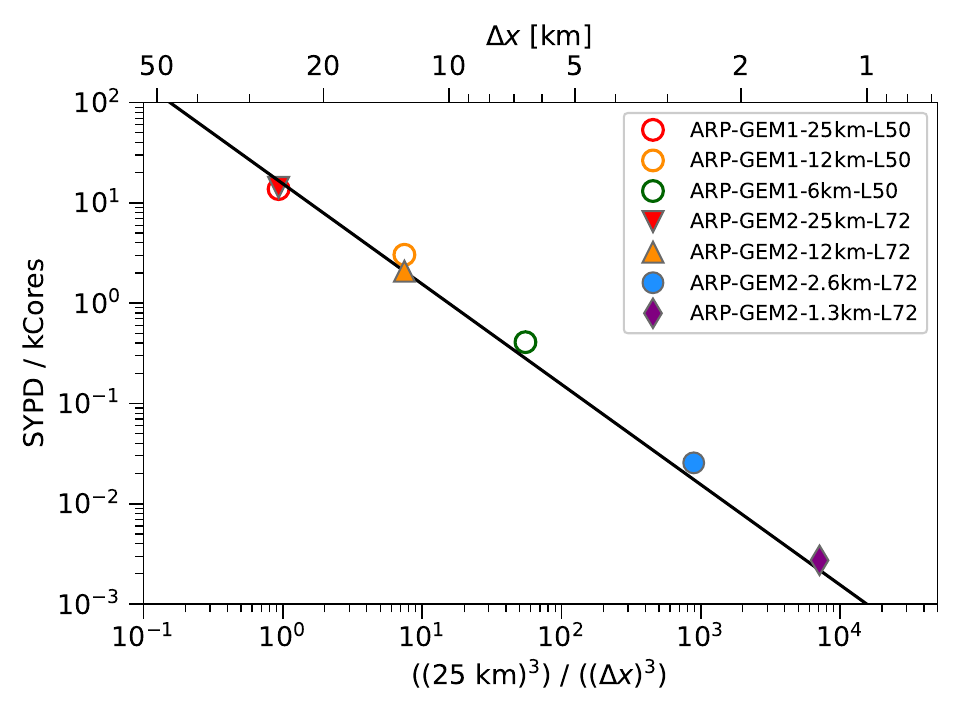}}
\caption{Simulated years per day (SYPD) normalized by the number of kilo-cores as a function of the cube of the inverse of the normalized grid spacing $(25/\Delta x)^3$ for ARP-GEM2 (72 vertical levels) configurations at 25 km (red), 12.6 km (orange), 6.3 km (green), 2.6 km (blue), and 1.3 km (violet) resolutions, and for ARP-GEM1 (50 vertical levels) configurations at resolutions from 25 km to 6.3 km (open circles). The line represents computational costs estimated by idealized scaling from the ARP-GEM2 25-km simulation.}
\label{fig:sypd}
\end{figure}

In GS25, the model demonstrated near-scalable performance for resolutions up to 6 km.
This scalability remains valid for the highest-resolution simulations at 2.6 km and 1.3 km (Figure \ref{fig:sypd} and Table \ref{tab:configurations}), as shown by computational efficiency scaling approximately linearly with 1/$(\Delta x)^3$, where $\Delta x$ is the horizontal grid spacing.
The factor $1/(\Delta x)^2$ reflects the number of horizontal grid columns, while the additional factor $1/\Delta x$ accounts for the scaling of the timestep with resolution. This scaling approximately follows from the Courant–Friedrichs–Lewy (CFL) stability condition \citep{courant-1928}.

Ideally, horizontal and vertical resolution should vary consistently \citep{lindzen-1989}.  Here, for practical reason, the vertical levels are fixed a priori and are not adjusted when switching from one resolution to another. Hence, the present scaling excludes vertical levels.

The model's scalability up to a 1.3 km resolution is an important result, given the high computational cost of kilometer-scale simulations. The 1.3 km resolution configuration performs 46 simulated days per day (SDPD) over 361 nodes on the Météo-France supercomputer Belenos\footnote{https://www.top500.org/system/179853. Each computational node is equipped with two AMD Epyc Rome processors, each with 64 cores operating at at 2.25 GHz.}.
Note that the coarsening factors in the 1.3 km configuration are larger than in other configurations. We estimate that, using the low-resolution coarsening factor, the cost would increase by at most 10–20 \%, which does not affect the conclusions on model scalability

At 2.6 km resolution, one-year simulations can be achieved at a reasonable cost of 1000 kh.CPU. This confirms the GS25 conclusion that centennial-scale climate simulations are feasible with ARP-GEM at resolutions on the order of 2 to 3 km.

\section{Simulations and resolution-dependent calibration}
\label{sec:simulations}

\subsection{Gradual suppression of deep convection}
\label{sec:diluted}

As resolution increases, the portion of the atmospheric flow that is explicitly resolved increases. To account for this, deep convection schemes are often made scale-aware through mass-flux closures. These closures apply an empirical rescaling of the cloud-base mass flux as a function of grid spacing, reducing the strength of parameterized convection as resolution increases \citep{arakawa-2011, grell-2014, kwon-2017, ecmwf-2019, freitas-2020, becker-2021}.
The reduction in convective activity with increasing resolution is represented by a decrease in mass flux toward zero. Other ways to conceptualize the diminishing role of parameterized convection at finer scales can be considered. 

A well-established concept in atmospheric modeling is that entrainment rates vary with the depth of convection: deeper clouds are associated with lower entrainment (and detrainment) rates \citep[e.g.][]{malkus-1959, simpson-1969, simpson-1971, ecmwf-2009}. This relationship is widely implemented in atmospheric models \citep{villalba-pradas-2022}.
Convection schemes based on a spectrum of cloud types are commonly used to represent convection \citep{arakawa-1974}. In Tiedtke-type schemes, the entire cloud population is represented by a single characteristic updraft. The difference in entrainment often accounts for the most significant distinction between deep and shallow convection schemes and can be the key factor in switching from one type of convection to another \citep{gregory-2001,bechtold-2008,ecmwf-2009,delgenio-2012}.

This dependence of entrainment on updraft size provides a framework for the gradual removal of deep convection parameterizations. As resolution increases, progressively smaller convective cells are gradually resolved by the model. 
The fraction of the entire cloud population or convective clusters represented by subgrid-scale convection must be considered increasingly shallow, with larger entrainment and detrainment rates.
Additionally, a significant feature is that simulations with the deep convection scheme turned off can be viewed as simulations with an infinitely large entrainment rate \citep{becker-2017}, which is consistent with this framework.
More precisely, within the context of the Tiedtke deep convection scheme, simulations without parameterized deep convection can be seen as having an entrainment rate large enough to prevent convection from being triggered, rather than requiring it to be infinite. 
Indeed, convection is activated when the cloud depth exceeds a given threshold (set to 300 hPa), which cannot be reached if the entrainment rate is sufficiently large. 

Finally, gradually increasing entrainment and, correspondingly, detrainment rates toward infinity as resolution increases provides a pathway toward the full suppression of parameterized convection.
This line of reasoning is particularly interesting because large entrainment rates or the imposition of a minimum entrainment rate generally improve simulations at low resolution, favoring variability and the MJO \citep{tokioka-1988, kim-2012}, decreasing the double ITCZ bias \citep{oueslati-2013}, and improving the representation of intermediate daily precipitation regimes \cite[e.g.][]{kooperman-2018}, even if not systematically. Indeed, a large entrainment tends to degrade the mean precipitation pattern \citep{kim-2012}. 

\begin{table*}[h!]
\caption{List of experiments. The top-of-atmosphere (TOA) outgoing LW radiative flux (OLR), net SW TOA radiative flux, surface latent heat flux (LHF) and sensible heat flux (SHF) are provided for a three-year period (2007–2009), except for the 1.3 km simulation, which is available only for the year 2007. The CERES radiative fluxes values are indicated. Model tuning parameters are detailed in the text.}
\begin{center}
\begin{tabular}{lccccccccccc}
 & OLR & net SW & $k_{\mathrm{au},i}$ & IF$_{\mathrm{sw}}$ & $\varepsilon_{\mathrm{up}}$ & $\delta_{\mathrm{up}}$ & RH$_{\mathrm{c},\mathrm{high}}$ & RH$_{\mathrm{c},\mathrm{low}}$ & $q_{l,0}$ & V$_\mathrm{sed}^\mathrm{solid}$ & \\
 & W m$^{-2}$ & W m$^{-2}$ & 10$^{-3}$ s$^{-1}$ & & 10$^{-3}$ m$^{-1}$ & 10$^{-4}$ m$^{-1}$ & & & mg kg$^{-1}$ & m s$^{-1}$ & \\
\hline
Obs, 2007 & 240.0 & 240.4 & & & & & & & & &  \\
\hline
1.3km & 239.3 & 238.2 & 1.05 & 0.90 & 2.8 & 2.3 & 0.60 & 0.90 & 1200 & 4.0 &  \\
DYAMOND3 1.3km &  &  &  1.03 & 0.79 & 2.9 & 2.3 & 0.60 & 0.895 & 1200 & 4.0 &  \\
\hline
Obs, 2007-2009 & 239.6 & 240.5 & & & & & & & & & \\
\hline
2.6km & 239.6  & 240.7 & 1.15 & 0.71 & 2.6 & 1.8 & 0.60 & 0.91 & 1000 & 2.0 &  \\
DYAMOND3 2.6km &  &  & 1.15 & 0.71 & 2.6 & 1.8 & 0.60 & 0.91 & 1000 & 2.0 &  \\
\hline
2.6km-nodeep & 244.9 & 243.3 & 1.15 & 0.71 &  $\infty$  &  $\infty$  & 0.60 & 0.91 & 1000 & 2.0 &  \\
\hline
2.6km-nodeep-tun & 240.5 & 242.1 & 0.66 & 0.63 &  $\infty$  &   $\infty$  & 0.60 & 0.91 & 1000 & 2.0 &  \\
\hline
2.6km-ed+ & 239.9 & 240.7 & 0.80 & 0.68 & 3.6 & 3.4 & 0.60 & 0.91 & 1000 & 2.0 &  \\  
\hline
12km & 239.8 & 240.1 & 1.10 & 0.71 & 2.6 & 1.8 & 0.70 & 0.94 & 350 & 0.9 &  \\
\hline
12km-nodeep & 245.3 & 239.9 & 1.10 & 0.71 & $\infty$ &  $\infty$  & 0.70 & 0.94 & 350 & 0.9 &  \\
\hline
25km & 239.8 & 240.5 & 1.25 & 0.71 & 2.6 & 1.8 & 0.80 & 0.97 & 300 & 0.9 &  \\
\hline
25km-nodeep & 245.4 & 237.7 & 1.25 & 0.71 &   $\infty$ &  $\infty$  & 0.80 & 0.97 & 300 & 0.9 & \\
\hline
25km-ed- & 239.8 & 241.4 & 1.60 & 0.71 & 1.8 & 0.75 & 0.80 & 0.97 & 300 & 0.9 &  \\
\hline
\end{tabular}
\label{tab:tuning}
\end{center}
\end{table*}

\subsection{Simulations and resolution-dependent calibration}
\label{sec:resoltun}

We perform two sets of simulations. First, we assess sensitivities to both the convection scheme and resolution using a suite of simulations ranging from 25 km to 2.6 km resolution, with and without an active deep convection scheme.
These simulations, along with their physical parameter differences, are summarized in Table \ref{tab:tuning}.
Second, we also document a 1.3 km simulation run for the year 2007, although it is not analyzed in detail in the present study. This simulation, in addition to the ARP-GEM2-2.6km simulation have served as the basis for the DYAMOND3 intercomparison exercise simulations covering January 2020 to February 2021 \citep{takasuka-2024b}.

The DYAMOND3 2.6 km and 1.3 km simulations are conducted with diluted deep convection, i.e., larger entrainment and detrainment rates. Larger values are used for the highest resolution, consistent with the reasoning in Section \ref{sec:simulations}\ref{sec:diluted}.
Parameters were selected as a trade-off, reflecting the need for some compromises, as shown below. Ultimately, the decisive factor was the accuracy of the OLR pattern representation.
Note that, in the DYAMOND3 simulation, the SW inhomogeneity factor (IF$_{\mathrm{sw}}$) has been revised to 0.79 to increase incoming SW radiation. Other parameter differences are minor. 

Sensitivity to the convection scheme is assessed at 2.6 km resolution. Corresponding simulations at lower resolutions, 12.6 km and 25 km, are also shown to examine the effect of resolution.
The low-resolution simulations use the same convective parameters as the 2.6 km configuration, corresponding to more diluted deep convection than in ARP-GEM1 (GS25).
These simulations are retuned from the 2.6 km configuration so that radiative balance is restored and mean cloud cover is roughly unchanged.

Simulation without parameterized deep convection are denoted by the subscript 'nodeep'. These simulations are conducted with identical values of parameters than simulation with the deep convection scheme activated. For the 2.6 km setup, an additional simulation without parameterized deep convection is carried out with retuning such that the radiative imbalance is closer to observed global mean values ('nodeep-tun').
Another 2.6 km simulation includes increased entrainment and detrainment rates (subscript 'ed$+$'), along with adjusted top-of-the-atmosphere (TOA) radiation. Finally, a 25 km simulation with reduced entrainment and detrainment (subscript 'ed$-$'), using parameter values consistent with those in ARP-GEM1, is included for comparison.

Calibration is performed in two steps, with parameter adjustments in the microphysics and grid-scale cloud schemes. First, parameter adjustments are made mainly to compensate for the sensitivity of cloud cover to resolution. Then, a final tuning is applied to bring the model closer to observed radiative imbalance. This process is carried out for time periods around year 2007.

The final tuning of LW radiation is performed with a highly uncertain parameter, the autoconversion rate of ice. Reducing autoconversion leads to more high-level cloud and decreases OLR. Unlike GS25, where both liquid and ice autoconversion were set to similar values for simplicity, they are now treated separately. The liquid autoconversion rate is fixed a priori and is not used in the final tuning of radiative balance.
Given the fixed SST-type configuration, SW radiation is less critical than in ocean-coupled experiments, which contrasts with LW radiation, tightly linked to precipitation \citep{pendergrass-2014}.
The SW component is finally adjusted using inhomogeneity factors for both ice and liquid water, parameters that rescale the optical depth of ice and liquid clouds. These factors are set to the same value for simplicity but could be adjusted separately for more precise tuning. Cloud properties have been pre-adjusted, ensuring that inhomogeneity factors remain within the range of 0.7 to 1 for the default configurations (i.e., configurations with parameterized deep convection).

In GS25, in addition to the inhomogeneity factors and autoconversion rates, the low-level cloud critical relative humidity RH$_{\mathrm{c},\mathrm{low}}$ was also adjusted to compensate for temporal and spatial resolution-dependent changes in cloud cover.
In the current study, a slightly larger set of parameters is adjusted across resolutions, providing more flexibility in correcting some resolution-dependent differences. We also consider the vertical velocity of solid precipitation V$_\mathrm{sed}^\mathrm{solid}$, the high level cloud critical relative humidity RH$_{\mathrm{c},\mathrm{high}}$, and the the liquid water autoconversion threshold $q_{l,0}$.

Low-level clouds tend to decrease at higher spatial and temporal resolutions, as observed in GS25, and this trend continues in simulations up to 1.3 km. Note that this behavior is counterintuitive, since at higher resolution the cloud scheme is generally expected to approach an all-or-nothing behavior (RH$_{\mathrm{c}}$ tending toward one). Possible explanations include the complexity of parameterized cloudy boundary-layer processes (turbulence and shallow convection), the constant number of vertical levels across resolutions, or non-monotonic resolution-dependent behavior.
As in GS25, the critical relative humidity at low levels RH$_{\mathrm{c},\mathrm{low}}$ is reduced with resolution to compensate for this decrease in low-level cloud cover. An increase in the width of the subgrid distribution of temperature and humidity leads to an increase in cloud fraction.
Additionally, the liquid water autoconversion threshold ($q_{l,0}$) is increased with resolution. 
This lower precipitation efficiency helps reduce the decrease in low-level clouds and likely compensates partly for the increase in in-cloud liquid water associated with the increase in RH$_{\mathrm{c},\mathrm{low}}$.    
Note that in the 25 and 12.6 km resolution configurations used in this study, the SW TOA radiation was ultimately tuned only based on these parameters to avoid too low values of the inhomogeneity factors.

The high level cloud critical relative humidity RH$_{\mathrm{c},\mathrm{high}}$ is reduced in high-resolution simulations to compensate for an increase in high cloud cover with resolution. 
Here, the sensitivity of cloud cover to the width of the subgrid distribution is opposite to that at low levels: an increase in RH$_{\mathrm{c},\mathrm{high}}$ (reflecting a larger width) leads to a decrease in cloud cover, as high-level grid points are saturated.
This adjustment does not strongly affect LW radiation, likely due to compensating effects between ice water content and cloud fraction.

Finally, the terminal velocity of solid precipitation is increased with resolution -- from 0.9 m s$^{-1}$ at low resolution to 4 m s$^{-1}$ at 1.3 km -- based on empirical results.
This is consistent with higher terminal velocities typical of convective precipitation (e.g., graupel and hail).
By reducing the time spent in the atmosphere, the increase in terminal velocity reduces accretion rates and evaporation of solid precipitation.
This increase helps reduce the positive bias in OLR in the tropics in simulations without deep convective parameterization (see Section \ref{sec:results}\ref{sec:olr}).
Larger values favor more extensive mid-level cloud cover, which may be related to the decrease in accretion rate. Low-resolution simulations cannot accommodate large values of solid precipitation terminal velocity, as mid-level cloud cover is higher at low resolution, making the simulation excessively reflective. In the 25 km and 12.6 km configurations, the vertical velocity of solid precipitation therefore needs to remain close to their original value (0.9 m s$^{-1}$), illustrating the relatively limited scope for adjustment.

\begin{figure*}
\centerline{\includegraphics[width=33pc]{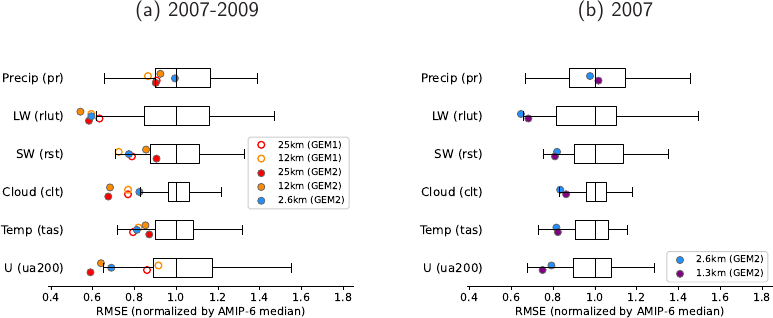}}
\caption{Annual normalized root-mean-square errors (RMSEs) in the climatology of precipitation (Precip), top-of-atmosphere longwave (LW) and net shortwave (SW) radiation, total cloud cover (Cloud), surface air temperature (Temp), and 200-hPa zonal wind (U), calculated against observational or reanalysis datasets. RMSE is normalized by the median value across 38 CMIP6 models (listed in Section 3 of the Supplementary Material of GS25). These median values for precipitation, LW radiation, SW radiation, total cloud cover, surface air temperature, and 200-hPa zonal wind are for 2007-2009 (respectively, 2007): 1.05 (1.20) mm~day$^{-1}$, 7.5 (8.5) W~m$^{-2}$, 10.9 (11.3) W~m$^{-2}$, 11.3 (11.6) \%, 2.4 (2.5) K, and 2.9 (3.2) m~s$^{-1}$. (a) RMSEs for ARP-GEM1 at 25-km (red circle) and 12-km (orange circle) resolutions, and ARP-GEM2 at 25-km (red dot), 12.6-km (orange dot), and 2.6-km (blue dot) resolutions, compared with the distribution of annual RMSEs for the 38 CMIP6 models during the period 2007–2009 (boxplot). (b) RMSEs for ARP-GEM2 at 2.6-km (blue dot) and 1.3-km (violet dot) resolutions for the year 2007, compared with the distribution of RMSEs for the 38 CMIP6 models (boxplot).}
\label{fig:rmse}
\end{figure*}

\section{Results}
\label{sec:results}

\subsection{Mean state errors}

Figure \ref{fig:rmse} compares the root mean square error (RMSE) of key climate variables for the ARP-GEM model with a large ensemble of 38 CMIP6 model versions, using the \textit{amip} experiment (see GS25). RMSEs are computed against climatologies from the Multi-Source Weighted-Ensemble Precipitation (MSWEP) dataset, version 1.2 \citep{beck-2017} for precipitation, from CERES-EBAF \citep{loeb-2009} for SW and LW TOA fluxes, from the CALIPSO-GOCCP product \citep{chepfer-2010} for cloud cover, from the BEST monthly dataset \citep{rohde-2013} for near-surface air temperature, and from ERA5 reanalysis data \citep{hersbach-2020} for the zonal wind at 200 hPa. All data are conservatively remapped onto a common 2.5$^{\circ}$ regular grid.

The main state variables are well captured compared to CMIP simulations across all resolutions, including the one at 1.3 km. 
These results indicate that simulations at the kilometer scale can be reasonably tuned.
This tuning is facilitated by the efficiency of ARP-GEM2 at high resolutions and by a multi-resolution tuning approach: the development of intermediate configurations allows for iterative tuning across resolutions and testing of the model's main parameters at lower resolution for further refinement

The TOA radiation fluxes are well represented in terms of mean biases (Table \ref{tab:tuning}) and spatial errors (Figure \ref{fig:rmse}). 
This reasonable representation of the radiation budget is associated with particularly accurate cloud cover compared to CMIP models. 
In particular, the low cloud cover is better represented through improvements in the mixing at the top of the boundary layer (Appendix \ref{app:topentr}).
However, low cloud cover tends to decrease at higher resolutions (not shown), although humidity is better represented. This behavior is in line with the sensitivity observed when increasing resolution from 50 km to 6 km (GS25). Additional vertical levels may be necessary to further improve low-level cloud representation at high resolution, along with enhancements in parameterization and further tuning.

The mean surface air temperature has slightly improved compared to ARP-GEM1. This is related to increased turbulence intensity, which is associated with a reduction in turbulent dissipation in this model version (Section \ref{sec:model}\ref{sec:physics}). Additionally, the surface temperature negative bias (GS25) tends to decrease as resolution increases, possibly due to changes in the strength of low-level mixing. The improvement in surface temperature may share a similar origin with the reduction in low-level clouds. A better representation of topography may also play a role \citep[e.g.][]{xu-2021}.  
The representation of zonal wind at 200 hPa is well captured in ARP-GEM2. The clear improvement from version 1 is due to the increase in vertical levels (GS25). 
From ARP-GEM1 to ARP-GEM2, the precipitation pattern does not show improvement at low resolution. Note that the ARP-GEM2 resolution simulations have larger entrainment and detrainment values, which are less favorable for the representation of this pattern, as shown for the 2.6 km configuration in Section \ref{sec:results}\ref{sec:olr}.

Finally, there is no apparent added value from higher resolution in representing precipitation patterns in these fixed-SST experiments. The errors tend to increase slightly with higher resolution. This may be due to insufficient model tuning, missing or inadequately represented processes, or to the absence of a notable resolution effect, with the current resolution still being too coarse. In particular, precipitation is governed by complex interactions between subgrid-scale and large-scale processes, meaning its representation depends heavily on parameterized processes that still require calibration.

\begin{figure*}[h]
 \centerline{\includegraphics[width=34pc]{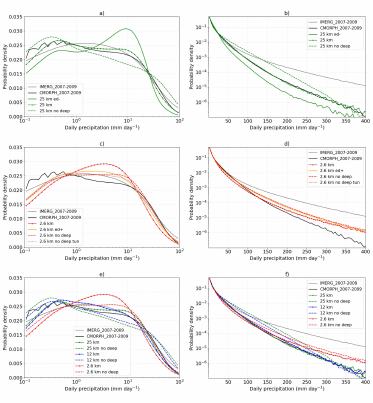}}
\caption{Probability density functions (PDF) of daily mean precipitation (in mm day$^{-1}$) over the tropical domain ($20^{\circ}$S-$20^{\circ}$N) for IMERG and CMORPH datasets and for ARP-GEM2 simulations at 25 km (top),  at 2.6 km (middle) and simulations with and without parameterized deep convection at 25, 12.6 and 2.6 km (bottom). Simulations are detailed in Table \ref{tab:tuning}. The period used is 2007-2009 for all datasets. Precipitation is conservatively interpolated to a to a 0.25$^{\circ}$ $\times$ 0.25$^{\circ}$ grid. Left panels show low precipitation rates in the range [$10^{-1}$-$10^{2}$] mm day$^{-1}$, uniformly binned on a $\log_{10}$ scale (50 bins). Right panels show high precipitation rates, binned with a size of 5 mm day$^{-1}$.}
\label{fig:pdf}
\end{figure*}  

\begin{figure*}[h]
 \centerline{\includegraphics[width=25pc]{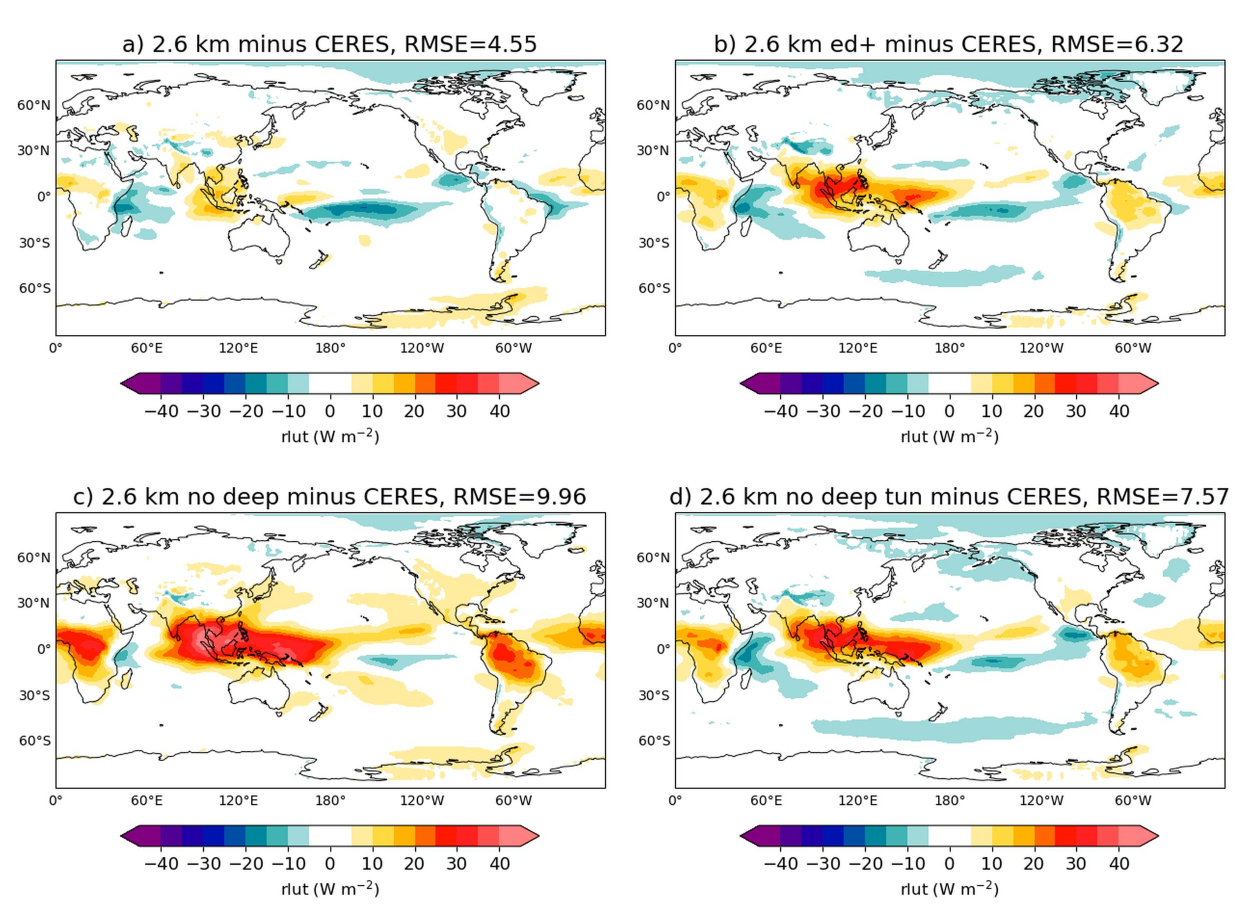}}
\caption{Annual mean of the OLR anomaly with respect to CERES observational dataset (period 2007-2009) for (a) ARP-GEM2-2.6km, (b) ARP-GEM2-2.6km-ed$+$, (c) ARP-GEM2-2.6km-nodeep, and (d) ARP-GEM2-2.6km-nodeep-tun.}
\label{fig:olr}
\end{figure*}

\subsection{Probability distribution of daily precipitation}
\label{sec:pdf}

Figure \ref{fig:pdf} shows the probability density function (PDF) of daily precipitation in the tropics (20$^\circ$S–20$^\circ$N) for simulations at 25 km, 12.6 km, and 2.6 km resolutions, with and without parameterized deep convection, and compares them with observational datasets: the Integrated Multi-satellite Retrievals for the Global Precipitation Measurement (IMERG) dataset, version 06 \citep{huffman-2019} and the Climate Prediction Center MORPHing technique (CMORPH) product, version 1.0 \citep{xie-2017}. All precipitation data are remapped to a 25 km grid. The results presented are for the years 2007–2009, with no sensitivity to the specific year used; any single year would suffice. The observed PDFs are characterized by large uncertainties, as shown by the differences between the observational datasets. As detailed in Section \ref{sec:simulations}, all simulations with the deep convection scheme activated use identical parameters, except for ARP-GEM2-2.6km-ed$+$, which has more dilute convection, and ARP-GEM2-25km-ed$-$, which uses less dilute convection.

At a 25 km resolution (Figs. \ref{fig:pdf}a-b), the low entrainment and detrainment values produce a precipitation PDF with a distinct peak around 10 mm day$^{-1}$. In contrast, high-precipitation regimes (above 30 mm day$^{-1}$) are associated with lower precipitation frequencies than observed (Fig. \ref{fig:pdf}b). 
This underestimation of extreme precipitation is similar to that seen in ARP-GEM version 1 (e.g. Fig. 16 in GS25). Both ARP-GEM1-25km and ARP-GEM2-25km-ed$-$ use the same entrainment and detrainment coefficients ($\varepsilon_{\mathrm{up}}$ = 1.8 $\cdot$ 10$^{-3}$ m$^{-1}$ and $\delta_{\mathrm{up}}$ = 0.75 $\cdot$ 10$^{-4}$ m$^{-1}$), resulting in similar PDFs despite differences in other model parameters and physics. 

As entrainment and detrainment increase, the peak in the intermediate precipitation regime decreases.
Without parameterized deep convection -- which represents the extreme case of this sensitivity, corresponding to infinite entrainment and detrainment  (Section \ref{sec:simulations}\ref{sec:diluted}) -- these effects become even more pronounced.
These results are consistent with large precipitation regimes being dominated by grid-scale precipitation from the microphysics, and intermediate regimes (around 10 mm day$^{-1}$) being dominated by convective precipitation, as in \citet{kooperman-2018}.

This behavior is not unique to this model but appears characteristic of climate models in terms of sensitivity to the entrainment rate \citep{kooperman-2018} or to the deep convection scheme being turned off at low \citep[e.g.][]{maher-2018} or high resolution, as seen with the Integrated Forecasting System (IFS) model \citep{becker-2021, takasuka-2026}. 
Convection schemes tend to concentrate rainfall into a predominant range. Indeed, most climate models show distinct peaks in their PDFs related to convective precipitation \citep{ahn-2024}.
This effect may be particularly strong in bulk schemes that represent all convection with a single mean updraft, such as the Tiedtke scheme, in contrast to spectral schemes.

Note that grid-scale precipitation can also produce distinct peaks in low-precipitation regimes \citep[e.g.][]{ahn-2024}. This is also observed in the very low range. This peak is likely related to grid-scale microphysics in low-level clouds, such as stratocumulus clouds. 

At low resolution and with large entrainment and detrainment rates, the peak in the intermediate precipitation regime decreases (Fig. \ref{fig:pdf}a), while the occurrence of larger precipitation amounts increases (Fig. \ref{fig:pdf}b), bringing the distribution closer to observations. These results point to an improved representation of the precipitation PDF when parameterized deep convection is more diluted. Without parameterized convection, the biases in the intermediate and heavy rain regimes may be reversed.

At high resolution, the sensitivities obtained at lower resolution remain valid. Increasing entrainment and detrainment rates allows a smooth transition between lower entrainment rates and convection turned off, as illustrated in Figs. \ref{fig:pdf}e-f.
However, the shape of the PDFs differ slightly.
With increasing resolution, simulations without parameterized deep convection tend to more closely match the observed distribution, with the intermediate regime being more fully covered by grid-scale precipitation (cf dashed lines in Fig. \ref{fig:pdf}c), accompanied by a decrease in the large precipitation regime (see dashed lines in Fig. \ref{fig:pdf}d). 
This suggests a decreasing need for parameterized convection to represent the daily precipitation PDF. 
More diluted convection appears to be more suitable as resolution increases.

\begin{figure*}[h]
 \centerline{\includegraphics[width=25pc]{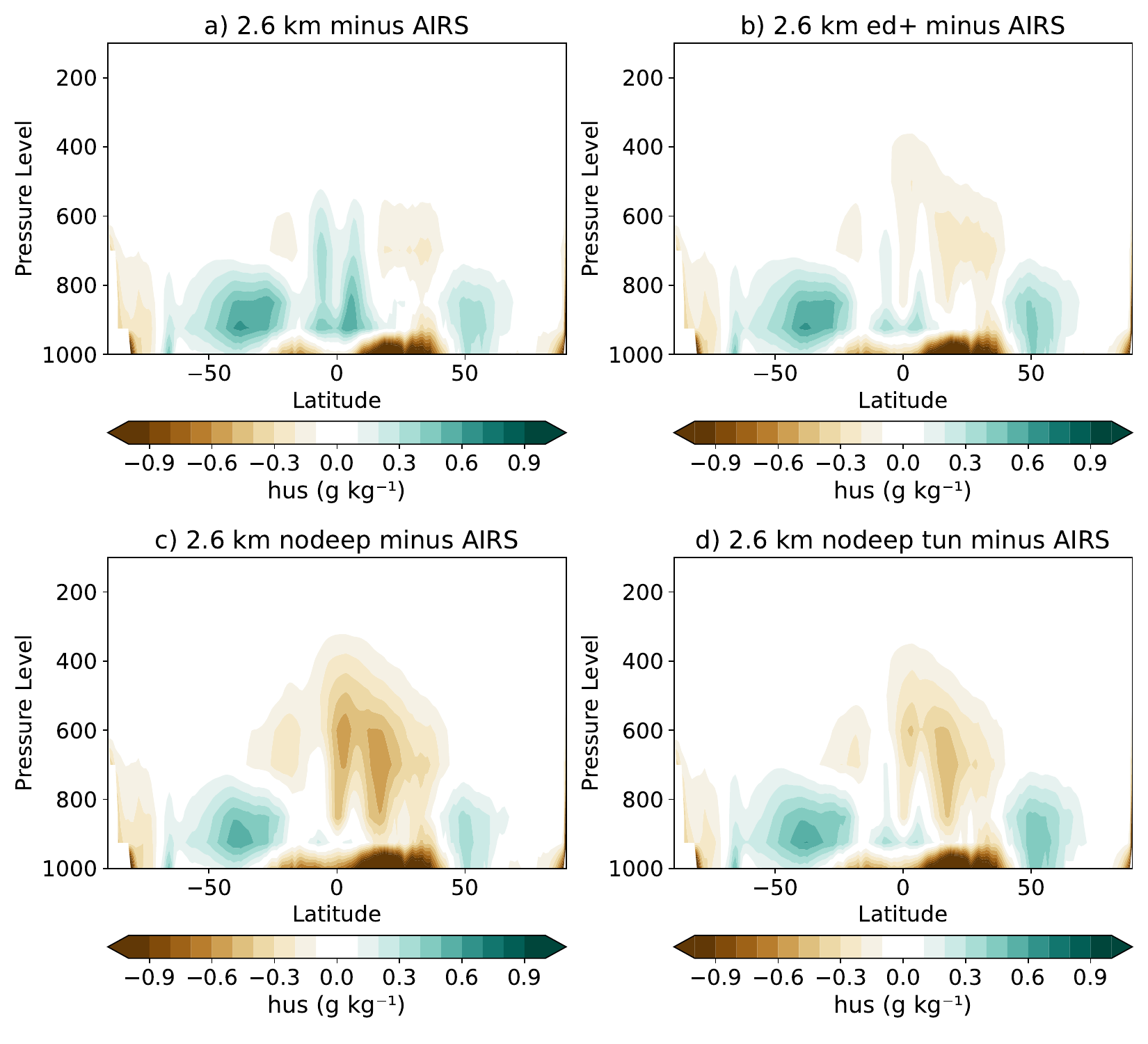}}
\caption{Annual zonal mean of the specific humidity anomaly with respect to AIRS reanlysis dataset (period 2007-2009) for (a) ARP-GEM2-2.6km, (b) ARP-GEM2-2.6km-ed$+$, (c) ARP-GEM2-2.6km-nodeep, and (d) ARP-GEM2-2.6km-nodeep-tun.}
\label{fig:hus}
\end{figure*}

\begin{figure*}[h]
 \centerline{\includegraphics[width=25pc]{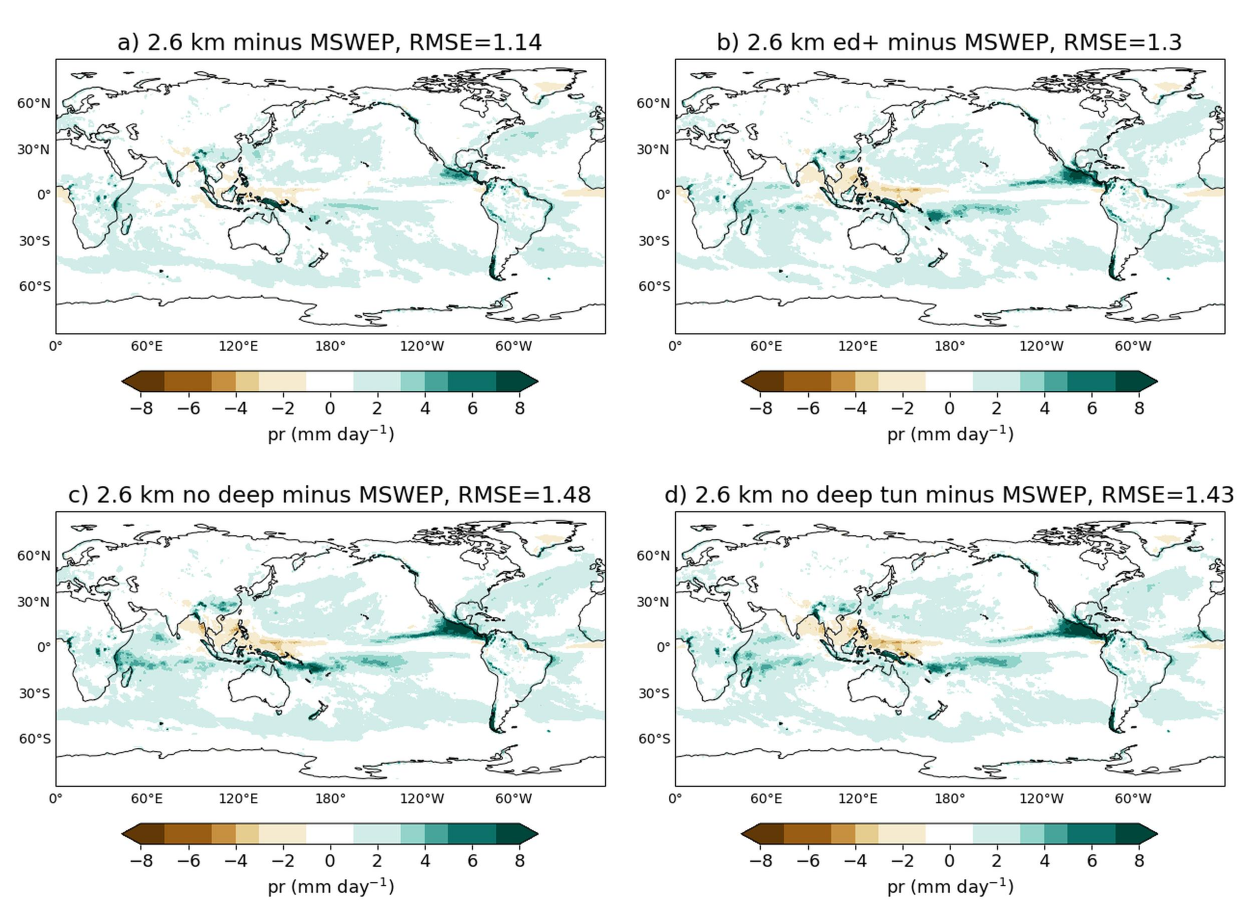}}
\caption{Annual mean of the precipitation anomaly with respect to MSWEP observational dataset (period 2007-2009) for (a) ARP-GEM2-2.6km, (b) ARP-GEM2-2.6km-ed$+$, (c) ARP-GEM2-2.6km-nodeep, and (d) ARP-GEM2-2.6km-nodeep-tun.}
\label{fig:pr}
\end{figure*}

Additionally, it is interesting to note that differences between simulations with and without deep convection tend to decrease slightly as resolution increases, particularly in the large-precipitation regimes, even though the deep convection scheme parameters remain unchanged across all model configurations. 
The importance of parameterized convection in representing precipitation seems to decrease in favor of large-scale processes.
\cite{wedi-2020} noticed no significant differences in the precipitation PDF across resolutions in their four-month 9 km and 1.4 km simulations. However, a closer look at our simulations suggests that, even if small, the differences between model resolution configurations are tangible. 

Finally, at low resolution, a lower entrainment rate appears necessary, whereas at high resolution, a higher entrainment rate -- or turning off the deep convection scheme -- is favored, consistent with the decreasing role of parameterized deep convection as resolution increases. 
The daily precipitation PDF is best represented at high resolution without parameterized convection. This improved representation without convection parameterization moderates results from some previous studies suggesting that parameterized convection favors the representation of precipitation PDFs \citep{becker-2021, takasuka-2026}.
However, removing the deep convection scheme can affect other aspects of model behavior, such as the mean state, as shown in the following sections.

\subsection{OLR, humidity and precipitation patterns}
\label{sec:olr}

Figures \ref{fig:olr} and \ref{fig:hus} show the mean OLR pattern and the zonal-mean specific humidity for simulations with and without parameterized deep convection. In simulations with the deep convection scheme turned off, the tropics are drier and exhibit higher OLR values. These biases are consistent: a drier tropical atmosphere is accompanied by a reduction in high-cloud amount (not shown), which in turn leads to increased OLR.
This effect is likely common to most models and is consistent with the drying \citep{maher-2018} and the decrease in high-level clouds and LW cloud radiative effect \citep{webb-2015} observed in low-resolution climate models when deep convection is turned off.
It shares similarities with simulations using the Icosahedral Nonhydrostatic model ICON, showing a drier troposphere at the equator and a moister 800–900 hPa layer in the extratropics \citep{kroll-2025}.
This behavior contrasts with the common view that parameterized deep convection tends to dry the atmosphere when considered in isolation. However, when the deep convection scheme is turned off, convective or ascending motions still occur due to the large-scale overturning circulation. These motions are more efficient at drying the atmosphere than when convection is also represented at finer scales by a parameterized convection scheme.

The change in OLR between simulations with and without parameterized deep convection is significantly greater in the tropics than in the extratropics, consistent with the stronger influence of convection in tropical regions. 
In particular, without parameterized deep convection, OLR is too large over continents, including the Maritime Continent.
Variations in the ice autoconversion rate ($k_{\mathrm{au},i}$) affect OLR more uniformly across regions.
The dry bias is still present after retuning of the global-mean TOA radiation in the 2.6km-nodeep-tun simulation, although its magnitude is reduced (Fig. \ref{fig:hus}d).
While the simulation is close to global radiative equilibrium (Table \ref{tab:tuning}), OLR remains too large in the deep tropics and exhibits an opposite bias in the extratropics (Fig. \ref{fig:olr}d).
The snow terminal velocity ($V_\mathrm{sed}^\mathrm{solid}$) has a predominant effect in the tropics, as mentioned earlier, but its influence is not sufficient to recover the LW biases observed when convection is turned off. 

The path toward fully turning off deep convection is not a straightforward function of the closure alone: decreasing the intensity of convection through the mass-flux closure can lead to a moistening effect (not shown). This suggests that the closure parameter is not appropriate for a smooth transition toward suppressing subgrid convection. In contrast, increasing the entrainment rate leads to intermediate biases in OLR and humidity, with patterns lying between those of a low entrainment rate and those observed when the deep convection scheme is turned off (Fig.~\ref{fig:olr}b and \ref{fig:hus}b), providing a more straightforward path toward turning off the deep convection parameterization.

Without parameterized convection or at very large entrainment rates, the precipitation pattern tends toward a double ITCZ structure (Fig.~\ref{fig:pr} and Fig. S3).
The precipitation pattern resembles that of the DYAMOND1 models, with large precipitation biases occurring over the ocean in the convergence zones of the Hadley–Walker circulation \citep[e.g. Fig. 2 in][]{schneider-2024}.
Other studies show that the double ITCZ bias may persist as resolution increases to at least 5 km \citep{kroll-2025}, including in km-scale ocean-coupled experiments \citep{segura-2022,segura-2025}.
The precipitation pattern also shares strong similarities with that obtained from low-resolution climate models \citep{maher-2018}, with wet biases in the eastern equatorial Pacific and Atlantic and dry biases over maritime continent and South Asia.
As entrainment decreases, the double ITCZ bias is reduced. 
This effect contrasts with the results of \citep{oueslati-2013}, in which larger entrainment instead reduces the double ITCZ structure in a low-resolution model. The improvement of the precipitation pattern when parameterized convection is included is fully consistent with what can be obtained at low resolution \citep[e.g.][]{kim-2011}.

The structure of the precipitation pattern and its sensitivity to the entrainment rate align with the behavior observed for OLR and humidity.
The removal of parameterized convection tend to produce an organized precipitation pattern embedded within the general circulation. This circulation–precipitation coupling is more efficient at removing tropospheric water than when parameterized convection is used. This is consistent with a positive feedback mechanism that strengthens the circulation’s ability to dry the troposphere, as precipitation becomes more concentrated and organized within it. Such a feedback is common in ITCZ dynamics and plays a key role in shaping the coupling between precipitation patterns and circulation \citep[e.g.][]{dixit-2018}.  

Parameterized convection tends to disperse precipitation and reduce maxima in convergence zones. This effect shares similarities with the cluster-scale sensitivity reported in \cite{takasuka-2026}, where convective clusters tend to be larger in spatial extent but weaker in precipitation rates when parameterized convection is used.
On seasonal to climatological scales, tropical precipitation is closely linked to local SST \citep[e.g.][]{good-2021}. In a simple fixed-SST framework, a deep convection scheme that captures the mean sensitivity to SST leads to an accurate representation of the climatological precipitation pattern.
Giving the fixed SST framework, deep convection pararameterization may be efficient at representing this sensitivity and redistribute, on average, precipitation more realiscally.

Finally, parameterized deep convection -- even when diluted with low entrainment rates -- allow the precipitation pattern, OLR, and humidity biases to improve. 
Further investigation is needed to determine whether better tuning or improvements in the representation of physical processes, such as changes in microphysics or turbulence, could achieve this balance without degrading other aspects of the simulated climate. Additionally, since the model uses the hydrostatic assumption, the role of non-hydrostatic effects should be examined, although some studies suggest they may not significantly affect results at such resolution \citep[e.g.][]{dueben-2020}. The effect of ocean coupling warrants further investigation. Finally, whether the convection scheme is still required at this resolution remains to be confirmed.

\begin{figure*}[h]
 \centerline{\includegraphics[width=50pc]{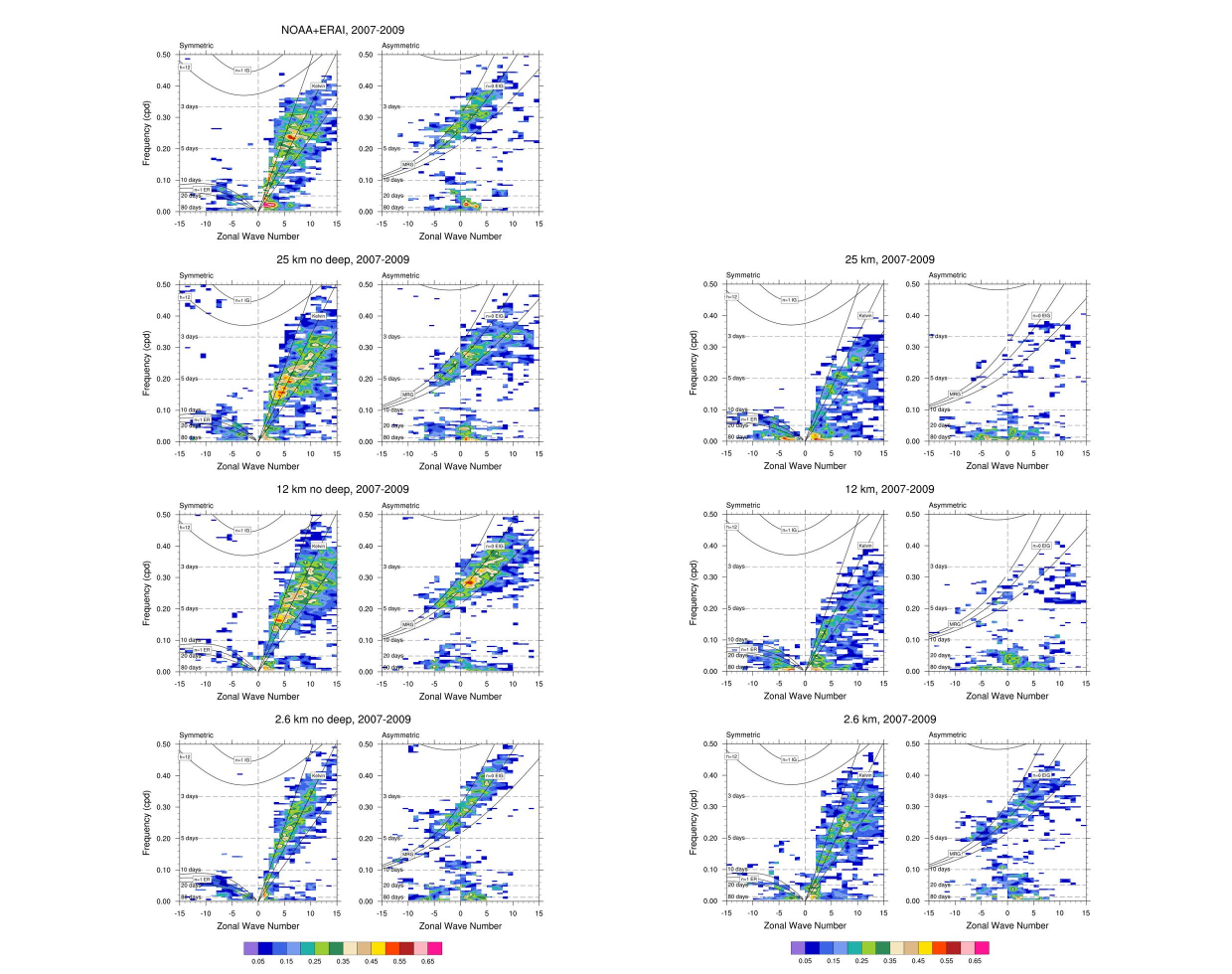}}
\caption{Frequency-wavenumber spectrum using daily outgoing LW radiation and zonal wind at 850 hPa for reanalysis (OLR from NOAA and wind from ERA-Interim) (upper panel) and for ARP-GEM2 at 25 km (second row), 12.6 km (third row), and 2.6 km (bottom) without (left) and with parametrized deep convection (right).}
\label{fig:spectres}
\end{figure*}

\subsection{Tropical Wave Variability}

Finally, we focus on tropical wave variability. Figure \ref{fig:spectres} presents the \cite{takayabu-1994} and \cite{wheeler-1999} spectra for the tropics (20$^\circ$S-20$^\circ$N). 
In the absence of parameterized deep convection, the simulation exhibits a clear spectral signature of equatorial waves, including Kelvin and mixed Rossby–gravity modes. 
However, when deep convective parameterization is included, these structures become substantially weaker (e.g., Kelvin waves) or even vanish, as in the case of mixed Rossby–gravity waves. This detrimental influence of the deep convection scheme is particularly evident in the 25 km and 12.6 km simulations. These results are consistent with those obtained at lower resolutions \citep{maher-2018}. They illustrate that deep convection parameterization is not always beneficial and can, in some cases, degrade or eliminate key climate features.

In the 2.6 km simulation without parameterized deep convection, the spectra are narrow, suggesting a more regular wave structure. The model may lack the perturbations typically introduced by small-scale convection, leading to a reduction in the complexity of variability representation. This behavior may be linked to the highly localized precipitation patterns observed in these simulations.

With parameterized convection included, the representation of waves improves as the resolution increases, even in simulations that use identical deep convection scheme parameters.
These results align with those reported by \cite{rackow-2025}.
The reduction in differences between simulations with and without parameterized convection mirrors that observed in the daily precipitation PDF (Section \ref{sec:results}\ref{sec:pdf}).
This improved variability representation suggests that errors associated with the deep convection scheme decrease with increasing resolution, or that convection is more embedded within the general circulation at higher resolutions. 

In all these simulations, the MJO signal is barely detectable. However, it can be observed, albeit weakly, in other simulations, mostly those with the deep convection scheme activated, indicating strong sensitivity to model parameters and variability. The absence of the MJO may be due to tuning issues, limitations in physical processes, lack of non-hydrostatic effects, or inadequate coupling with the ocean. 
In particular, the microphysics is expected to have an important impact on the MJO representation \citep{takasuka-2024a}. The ocean coupling can also play an important role in improving the MJO representation \citep[e.g.][]{demott-2019, savarin-2022, tseng-2022}.

\section{Conclusion}

This paper documents the version 2 of the ARP-GEM (Global Efficient and Multiresolution) atmospheric model, along with a set of kilometer-scale simulations exploring sensitivities to the representation of subgrid deep convection.
Version 2 of the model includes additional developments over ARP-GEM version 1 (described in detail in GS25), with modifications to both the physical and dynamical components (Section \ref{sec:model} and Appendix). In particular, these improvements aim to enable simulations at kilometer scales. 

The model’s scalability appears to be preserved at kilometer-scale resolutions (Fig. \ref{fig:sypd}). The ARP-GEM2 model is currently able of performing global simulations at 1.3 km horizontal resolution with 46 SDPD with about 46,000 CPUs (Section \ref{sec:model}\ref{sec:performances}). From a practical perspective, the ARP-GEM high computational efficiency enables short-term climate simulations, ranging from decadal to centennial scales, at a resolution of 2 to 3 km, with a reasonable amount of computational resources -- roughly 100 years with 100,000 kh.CPUs. These results demonstrate that centennial scale global simulations, such as those performed for climate services within regional modeling, can be conducted at these very high resolutions. The global approach helps minimize development efforts and reduce errors due to regional modeling, avoiding the need for nested downscaling.

Radiation is an expensive component. Its cost is reduced in ARP-GEM through the use of both temporal and spatial coarsening (also used for surface fluxes). The present radiation timestep and spatial coarsening factor for radiation and surface in the kilometer-resolution configurations are relatively large. Sensitivity tests, in which the radiation timestep is reduced from 20 min to the model timestep (4 min) and the spatial coarsening factor is reduced from 5.8 to 3 in the 2.6 km configuration, show no significant sensitivity (Supplementary Material).

The contribution of the deep convection scheme must be reduced when moving toward higher resolution. 
Indeed, as shown in Figs. \ref{fig:pdf}-\ref{fig:pr}, increasing the entrainment rate produces structures that lie between those of less diluted convection and those with deep convection turned off.
Consistent with theoretical and physical considerations, increasing entrainment (and detrainment) rates to make parameterized convection more diluted is a suitable approach for transitioning toward a configuration with spatial resolution sufficiently high to allow the deep convection parameterization to be turned off.
The shallow convection scheme was also made more diluted and less intense compared to ARP-GEM version 1. However, it was fixed a priori and not retuned depending on model resolution for the simulations analyzed in this study.

Special care was taken in model calibration. 
The model shows good performance in comparison with CMIP models, accurately representing the climatological patterns of the main climate variables (Fig. \ref{fig:rmse}). The high computational efficiency of the ARP-GEM model, along with the development of a suite of gradually varying resolution configurations, enables effective model tuning. In particular, it allows for managing TOA radiation close to the Earth-observed radiative balance.  
This suggests that simulations at 2.6 and 1.3 km resolutions can be reasonably tuned with this model. 

We perform sensitivity tests on the convection scheme at a 2.6 km resolution. They yield contrasting results. 
Variability tends to improve at high resolution and without parameterized convection.
With increasing resolution, the PDF of the daily precipitation is better represented without parameterized deep convection.
High-resolution simulated PDFs tend to closely match observed PDFs.
In particular, precipitation increases in the intermediate regime and decreases in the high regime, helping to correct biases that require parameterized deep convection (Section \ref{fig:pdf}). 
This suggests a decreasing role for convection parameterization. 
Wave power spectra appear more accurately represented without parameterized convection (Fig. \ref{fig:spectres}).
In addition, some differences between simulations with and without parameterized deep convection -- especially those related to tropical variability -- are reduced at high resolution, as shown in the daily precipitation PDFs (Section \ref{sec:pdf}) and tropical wave spectra. This may suggest a decreasing role of parameterized convection at high resolution or an improved coupling with the circulation.

In contrast, the mean state does not show such improvement with high resolution, and the main model structures are entrenched in larger biases when deep convection is turned off. Indeed, without parameterized convection, the pattern organization favors precipitation in concentrated, ascending regions, with a tendency toward a double ITCZ. 
The large-scale convective circulation efficiently dries the troposphere by enhancing precipitation efficiency, likely reinforced by a positive feedback. 
With parameterized deep convection, part of these biases are recovered, possibly by dispersing the convection, which prevents the concentration of ascent in privileged areas.  
Convection schemes should correct the model sufficiently without overcorrecting or introducing distortions. In particular, they must not reverse the sign of precipitation pattern biases present in the simulation without parameterized convection.

At the kilometer scale, relatively dilute parameterized deep convection appears helpful in reducing certain mean-state model biases. 
The apparent need for parameterized deep convection may align with the model's effective resolution, which remains large at this scale, with a multiplicative factor of 4 to 10 \citep{Abdalla-2013, ricard-2013}.
Further investigation is needed to determine whether other processes, such as changes in microphysics or the use of a non-hydrostatic core, could achieve improvements in both variability and mean state.
Future work will focus on extending sensitivity to microphysics and turbulence, and exploring the role of ocean coupling and non-hydrostatic effects.

\acknowledgments
We thank the three anonymous reviewers for their comments that helped to improve and clarify the manuscript. We also thank Gilles Bellon for discussions.

\datastatement

The DYAMOND3 simulations will be available through the Tokyo Node, which collects DYAMOND3 data. 
All CMIP6 model outputs are available via the portal: https://esgf-node.llnl.gov/search/cmip6. 
CERES-EBAF\_L3B\_Ed2-8 data (Table \ref{tab:tuning}, Figs \ref{fig:rmse} and \ref{fig:olr}) were obtained from https://ceres.larc.nasa.gov. The MSWEP\_V1.2 dataset (Figs \ref{fig:rmse} and \ref{fig:pr}) is available at www.gloh2o.org. IMERG\_V06 data (Fig. \ref{fig:pdf}) and AIRS data (Fig. \ref{fig:hus}) were obtained from https://disc.gsfc.nasa.gov. BEST temperature data (Fig. \ref{fig:rmse}) are available at https://berkeleyearth.org/data,  ERA5 data (Fig. \ref{fig:rmse}) and ERAI data (Fig. \ref{fig:spectres}) from https://cds.climate.copernicus.eu. CMORPH\_V1.0 data (Fig. \ref{fig:pdf}) were obtained from https://www.ncei.noaa.gov/data/cmorph-high-resolution-global-precipitation-estimates. Daily NOAA OLR data (Fig. \ref{fig:spectres}) are available at https://archive.data.noaa.gov.
The SURFEX code is available under a CECILL‐C License at the SURFEX website (http://www.umrcnrm.fr/surfex). XIOS can be downloaded from the XIOS website (https://forge.ipsl.jussieu.fr/ioserver).

\appendix[] 
\appendixtitle{Updates in ARP-GEM2 physics}

\section{Deep Convection}
\label{app:deep}

In ARP-GEM1, the entrainment formulation of the deep convection scheme is based on \citet{bechtold-2008} and \cite{ecmwf-2009}. The entrainment depends on relative humidity and a vertical scaling function based on the ratio of saturation specific humidity at the considered height to that at cloud base. In the triggering scheme, the test parcel updraft entrainment also depends on saturation specific humidity but is normalized by saturation specific humidity at the surface rather than at cloud base \citep{ecmwf-2009}.

To avoid such an ad hoc term in ARP-GEM2, we use in the triggering scheme the original formulation of \citet{jakob-2003} i.e., with an entrainment rate equal to $c_e/z$, with $c_e = 0.55$ and bounded to a minimum value $\varepsilon_{\mathrm{min}}$ (set to zero in this model version).
Note that the \citet{bechtold-2008} formulation remains in use for the final convective updraft. 
We have also implemented an optional entrainment parameterization consistent with that of the shallow convection scheme (depending on buoyancy over the square of vertical velocity) that can be used alternatively.

Additionally, the intensity of convection at cloud base is decreased by increasing the multiplicative mass-flux closure coefficient, $k_{\mathrm{cv}}$, from 1.35 to 2 (see GS25).
Finally, the deep convection autoconversion rate coefficient, $c_{00}$, is reduced to 1.$\cdot$10$^{-3}$s$^{-1}$.

\section{Cloud Top Entrainment and Penetrative Shallow Convection}
\label{app:topentr}

In ARP-GEM1, mixing at the top of the cloudy boundary layer is addressed through two separate representations (GS25): the mixing associated with penetrative shallow convection and the inclusion of a radiatively driven turbulent entrainment, expressed via a diffusion-like process \citep[][GS25]{lock-1998, ecmwf-2019}. The turbulent entrainment is applied only over the cloudy fraction, by scaling the turbulent fluxes by the cloud fraction. In contrast, shallow convection mixing is limited in cases where the convective overshoot exceeds a given distance but is applied over the entire domain area

To enhance consistency between the shallow convection and turbulence schemes in ARP-GEM2, the mass flux is set to zero over the cloudy fraction area at the top of the convective updraft. This adjustment is achieved by multiplying the shallow convection fluxes at the upper interface level $k - 1/2$ of layer $k$ by $1-\max(C_k, C_{k+1})$, where $C_k$ is the cloud fraction at level k.
This reduces mixing at the inversion and consequently increases cloud cover in stratocumulus regions. As a result, the limitation on overshoot depth (see GS25) can be slightly relaxed. This depth is still maintained at a non-zero value (20 m); otherwise, cloud cover dramatically decreases in stratocumulus regions.

\section{Shallow Convection}
\label{app:shall}

In the new version, shallow cloud depths are bounded by a maximum height value \citep[e.g.][]{deng-2003,ecmwf-2009}. Above this depth, no shallow cloud is allowed. This is consistent with the triggering of deep convection, which allows deep convection if clouds are thicker than a minimal depth (set to 300 hPa in this model version). Here, the shallow convection mass flux is linearly smoothed between $h_{max}^-$ and $h_{max}^+$, with $h_{max}^-=2.5$ km and $h_{max}^+ = 3.5$ km.

The shallow convective closure relates the convective fraction $\alpha_{u}$ at first flux level to the boundary layer depth $z_i$ (see details in GS25):
\begin{equation}
\alpha_{u}(z_{K-1/2}) = \min(\mathcal{C}_M z_{i}^{1/3},\alpha_{u,\text{max}})
\end{equation}
To prevent excessive moistening due to a reduction in shallow cloud precipitation (see next paragraph) and to limit overly active shallow convection, the convective area is reduced, by decreasing the scaling coefficient $\mathcal{C}_M$ from $0.015$ m$^{-1/3}$ to $0.010$ m$^{-1/3}$ and by limiting the maximum allowed area to 0.10 (compared to 0.30 in ARP-GEM1). 

In ARP-GEM1, shallow precipitation was diagnosed, and the liquid and ice autoconversion rates were likely too high, contributing to the positive precipitation bias observed over land (GS25). The simplest representation of shallow convection precipitation is to suppress it entirely. However, the precipitation pattern appears to improve when non-zero shallow precipitation is allowed. To simplify the model and reduce the number of parameters, shallow cloud precipitation is now treated through the grid-scale microphysics scheme, as done in other models: grid-scale precipitation is computed after including the contribution of the shallow convection scheme to the total cloud cover and water content. 

In ARP-GEM1, shallow clouds were likely too reflective, with large liquid water paths.
To ensure a smoother transition between the updraft in-cloud water and the environmental cloud water, the grid-box mean shallow convective cloud water $q_{c}^{\mathrm{sh}}$ is represented as follows:
\begin{equation}
q_{c}^{\mathrm{sh}} = \alpha_{\mathrm{up}}q_{c}^{\mathrm{up}} + (C^{\mathrm{sh}}-\alpha_{\mathrm{up}}) (q_{c}^{\mathrm{up}}+q_{c}^{\mathrm{env}})/2
\end{equation}
where $\alpha_{\mathrm{up}}$ is the convective fraction, and $C^{\mathrm{sh}} = k_\mathrm{cld} \alpha_{\mathrm{up}}$ is the shallow cloud fraction, with $k_\mathrm{cld}$ set to 2 (compared to 2.4 in ARP-GEM1).

The entrainment rate was previously modeled using a $B_{\mathrm{up}} / w_{\mathrm{up}}^2$ dependency \citep{fox-1970, gregory-2001} and bounded to a minimal value. In the current version, it is represented as the sum of a constant term and a $B_{\mathrm{up}} / w_{\mathrm{up}}^2$ term:
\begin{equation}
\varepsilon = C_o B_{\mathrm{up}} / w_{\mathrm{up}}^2 + \varepsilon_t
\end{equation}
with $C_o = 0.21$ and $\varepsilon_t = $ 0.0005 m$^{-1}$.
This can be interpreted as the sum of a turbulent entrainment rate and an organized entrainment rate \citep[e.g.][]{de_rooy-2013}.

A minimum detrainment rate, set to a relatively high value of 0.0015 m $s^{-1}$ for moist updrafts, was initially introduced to reduce precipitation over land (by reducing convective mass flux). This reduction in land precipitation is now likely also associated with other subsequent changes in the scheme, which, if implemented independently, produce similar results, such as adjustments to shallow precipitation intensity and a reduction in mass flux. The dry fractional detrainment rate $\delta_{d}$ is set to a constant value of 0.0008 m$^{-1}$.

The initial updraft properties are bounded, with reduced limits of 0.2 K for temperature and 100 g kg$^{-1}$ for humidity (instead of 1 K and 500 g kg$^{-1}$ in ARP-GEM1). 

We represent the momentum transport by shallow convection using the same theoretical framework as for thermodynamic variables \citep[e.g.][]{pergaud-2009}. The inclusion of this effect may influence wind patterns, such as jet streams. In some parameterizations, an additional term is introduced to account for the counteracting effect of the pressure gradient on momentum transport \citep[e.g.][]{gregory-1997, pergaud-2009}. Here, we simply introduce a scaling parameter to optionally reduce the intensity of the turbulent momentum flux. Nevertheless, no rescaling is applied in ARP-GEM2, as the parameter is set to one.

Under the hypothesis of an infinitesimally small updraft fraction (even though it is limited here), turbulent kinetic energy (TKE) is detrained from convective updrafts to the environment -- assuming that environmental TKE represents the entire grid-box area -- thereby adding a source term to the TKE budget: 
\begin{equation}
\frac{d\mathrm{TKE}}{dt} = \delta M_{\mathrm{up}} \frac{1}{2} w_{\mathrm{up}}^2
\end{equation} 
where $\delta$ is the fractional detrainment rate. This corresponds to the entrainment production term in the TKE equation for the environment in \cite{cohen-2020}. It was not found to have a strong effect on the model results, although it appears to increase mixing in the convective inhibition (CIN) region.
The transport of environmental TKE by the convective updraft can also be optionally included, but it does not have a significant impact.

\section{Turbulence}
\label{app:turb}

The intensity of turbulence is increased. More precisely, turbulent dissipation is reduced: the coefficient $C_\varepsilon$ that scales the dissipation term in the TKE equation is decreased from 1/1.18 to 1/1.4. In addition, the eddy diffusivity coefficient $K_e$ for vertical turbulent transport of TKE is decreased. It is expressed as $K_e = \alpha_e K_m$, where $K_m$ is the momentum eddy diffusivity, with $\alpha_e$ reduced from 2.7 to 2.

\section{Microphysics}
\label{app:micro}

The cloud-to-rain autoconversion coefficient, $k_{\mathrm{au},l}$ (see GS25), is decreased to $4\cdot10^{-4} s^{-1}$.
This contributes to an increase in low cloud amount, which tends to be reduced with increasing resolution. The intercept parameter in the snow size distribution, $N_{0s}$, and that in the liquid particle size distribution, $N_{0r}$, are both set to $8 \cdot 10^6$ m$^{-4}$.

\section{Dynamics and technical developments for kilometer resolution}
\label{app:kmscale}

A few modifications have been introduced to the dynamical and numerical parameter settings. In particular, the number of iterations in the scheme used to compute the departure points in the semi-Lagrangian advection has been increased from three to five \citep{diamantakis-2016}.

The procedure for generating orographic forcing files has been modularized, optimized, and parallelized to improve computational efficiency and scalability. In addition, portions of the code that produce the initial atmospheric and surface conditions have been revised and parallelized to accommodate the considerable increase in grid size associated with high-resolution configurations, reaching approximately 240 million grid points at 1.3 km resolution. These improvements are closely related to the evolution of the FA (Fichier ARPEGE) format library \citep{clochard-2002} which has been updated to effectively manage high-resolution datasets.

For such configurations, the hybrid MPI/OpenMP parallelization of ARPEGE/IFS (see details in GS25) is employed to optimize memory usage and computational performance. Further adjustments -- primarily related to the XIOS input/output software \citep{meurdesoif-2017} -- have been introduced to ensure the full effectiveness of OpenMP parallelization in all model configurations. As a result, the ARPEGE/XIOS interface has been redesigned and upgraded to fully support these capabilities.

\bibliographystyle{ametsocV6}

\end{document}